\documentclass[aps,prl,twocolumn,superscriptaddress,showpacs,showkeys]{revtex4}
\usepackage{graphicx,amsmath,amssymb,bm}

\newcommand{\be}{\begin{equation}}
\newcommand{\ee}{\end{equation}}
\newcommand{\g}{\, \text{G}}
\newcommand{\mt}{\, \text{mT}}
\newcommand{\khz}{\, \text{kHz}}

\newcommand{\la}{\Lambda}
\newcommand{\nmax}{N_{\rm max}}
\begin{document}

\title{Three-fermion problems in optical lattices}

\author{T.~Luu}
\email[E-mail:~]{tluu@llnl.gov}
\affiliation{N Division, Lawrence Livermore National Laboratory,
Livermore, CA 94551}
\author{A.~Schwenk}
\email[E-mail:~]{schwenk@u.washington.edu}
\affiliation{Department of Physics, University of Washington,
Seattle, WA 98195}

%\date{\today}

\begin{abstract}
We present exact results for the spectra of three fermionic atoms 
in a single well of an optical lattice. For the three lowest
hyperfine states of $^6$Li atoms, we find a Borromean state
across the region of the distinct pairwise Feshbach resonances. 
For $^{40}$K atoms, nearby Feshbach resonances are known for two 
of the pairs, and a bound three-body state develops towards 
the positive scattering-length side. In addition, 
we study the sensitivity 
of our results to atomic details. The predicted few-body phenomena 
can be realized in optical lattices in the limit of low tunneling.
\end{abstract}


\maketitle

{\it Introduction.--} Experiments with cold atomic gases make
it possible to study strong-interaction physics in a controlled
manner. When an atomic gas is loaded into an optical lattice,
typically a few atoms reside in each well. Therefore, optical lattices
can be used to investigate few-body phenomena when the tunneling 
barrier between potential wells is high~\cite{Jaksch}. In
dilute gases, the interactions are governed by the S-wave scattering 
length $a$, which can be tuned across atomic Feshbach resonances.
Consequently, three-fermion problems in optical lattices can
access nearly the entire landscape of fascinating few-body
phenomena. When all scattering lengths are large, the few-body
physics of dilute gases exhibits universal properties because 
there are no length scales associated with the interaction.
These universal aspects stretch across physics: For example,
the large scattering-length physics predicts a linear correlation
for ground-state energies of few-helium clusters or light
nuclei~\cite{BH}.

The first step towards realizing isolated few-atom systems 
was the formation of molecules from fermionic atoms in an
optical lattice~\cite{Michael}. In addition, the ground-state 
energy of two particles in a single well of an optical lattice 
was measured across a Feshbach resonance by rf dissociation 
to noninteracting fragments. The measured
ground-state energies (for varying scattering length) agree
very well with the theoretical prediction for two particles
in a harmonic oscillator potential~\cite{Michael}.

For three identical bosons, the scattering lengths of all
pairs can be tuned with one Feshbach resonance and the
three-boson problem is universal: On resonance, there is
an infinite tower of Efimov trimer states with consecutive 
binding energies $E_n/E_{n+1} \approx 515$ in free space~\cite{Efimov}. 
Efimov states become bound for a finite, negative scattering
length, and lead to resonances in the rate for three-body 
recombination. Recently, the first Efimov resonance was
observed in a gas of cold C{\ae}sium bosons~\cite{Kraemer}.
For three particles in a harmonic well, the spatial
confinement restricts the size of the Efimov states,
and thus the accumulation of bound states does not persist.
However, Stoll and K\"ohler have shown that it is possible to
study the first Efimov state isolated in a single well
of an optical lattice~\cite{SK}. This three-boson problem
was also studied by Jonsell {\it et 
al.}~in an adiabatic approximation~\cite{JHP}. 
Efimov trimers can be examples of Borromean systems
(when the scattering length is negative):
three-body bound states for which all pairs
are unbound. Other Borromean
systems are the $^6$He nucleus and the $^{11}$Li
halo nucleus ($^4$He or $^9$Li and two neutrons)~\cite{Zhukov}.

For fermions, in contrast to identical bosons, the Pauli
principle restricts S-wave interactions to
different hyperfine states, which have distinct pairwise
Feshbach resonances. In this Letter, we calculate exactly
the spectra of the three lowest hyperfine states of $^6$Li 
or $^{40}$K atoms confined to a single harmonic well of 
an optical lattice, using effective field-theory
interactions. For $^6$Li atoms, there are Feshbach 
resonances between all pairs~\cite{Li}, two of which overlap 
closely, and we find a Borromean state that extends across all
Feshbach resonances. For $^{40}$K atoms, Feshbach resonances
exist between two pairs, whereas the third pair interacts weakly.
In this case, we find that a three-body state only develops 
towards the positive scattering-length side, and is bound
around the doubly-interacting particle. The spectra for
three $^6$Li or $^{40}$K atoms in optical lattices are 
very rich, and the Bloch-Horowitz method employed here is ideally suited
to identify the angular momenta of the states. The predicted 
few-body phenomena can be realized in optical lattices in the 
limit of low tunneling.

\begin{figure}[t]
\begin{center}
\includegraphics[scale=0.45,clip=]{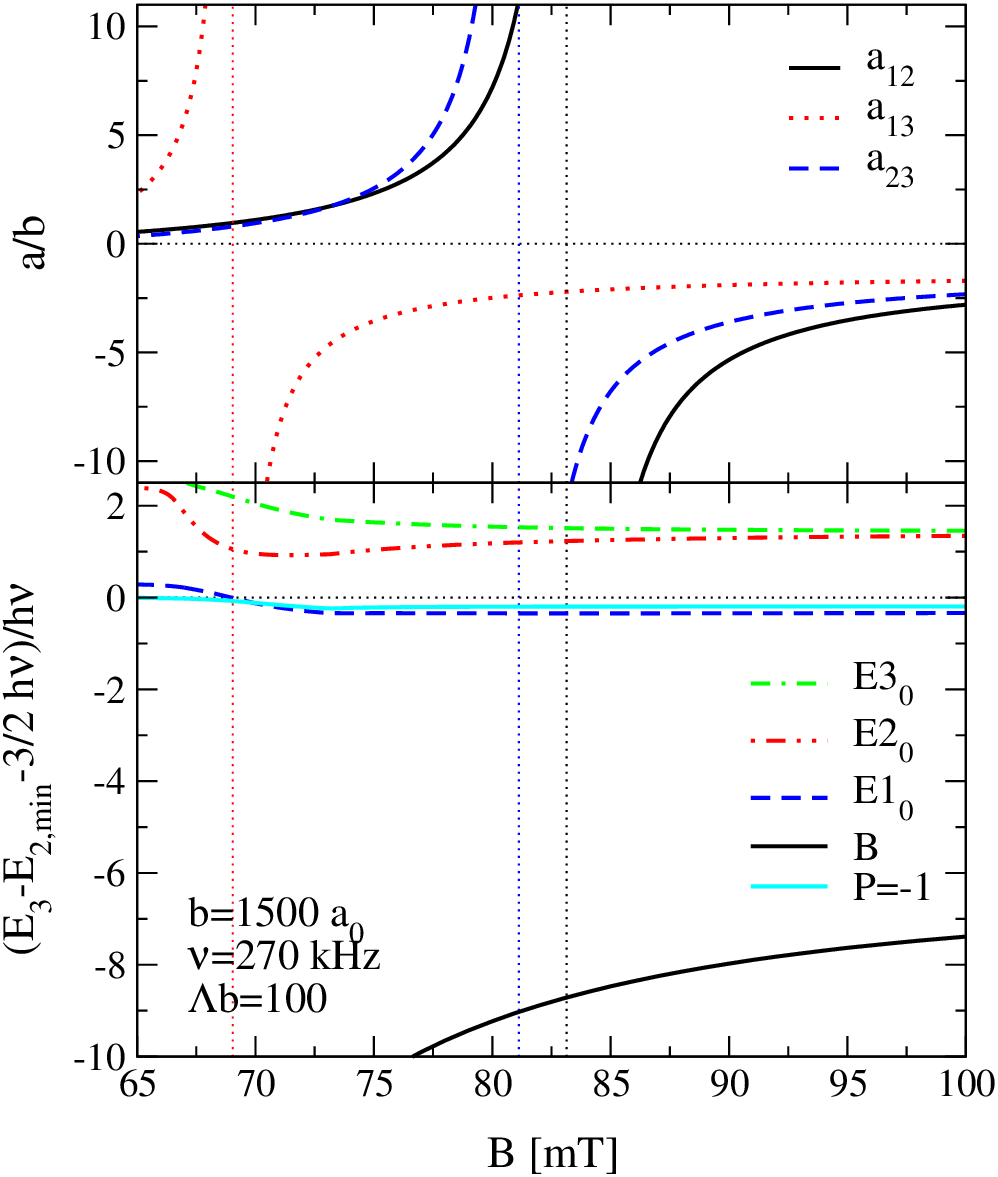}
\caption{Top: Feshbach resonances between the three lowest 
hyperfine states of $^6$Li atoms~\cite{Li}. Bottom: The
spectrum for the above $^6$Li states in a single 
well of an optical lattice with $\nu = 270 \khz$ versus
magnetic field. These BH results are for $\la b = 100$.
The three-body energy $E_3$ is measured from the three-body
dissociation threshold given by the minimal two-body energy
$E_{2,{\rm min}}$ (from Eq.~(\ref{elam})) of the pairs 
and the additional particle
in the noninteracting ground state $E_1 = 3/2 \hbar \omega$
($5/2 \hbar \omega$ for the negative parity state P=-1).
For $B > (<) 73.2 \mt$, $E_{2,{\rm min}}$ is determined from
$a_{12} (a_{23})$.
The state labeled B is the Borromean state, and En$_L$
denotes the n-th excited state with angular momentum $L$.
\label{Li6}}
\end{center}
\end{figure}

\begin{figure}[t]
\begin{center}
\includegraphics[scale=0.45,clip=]{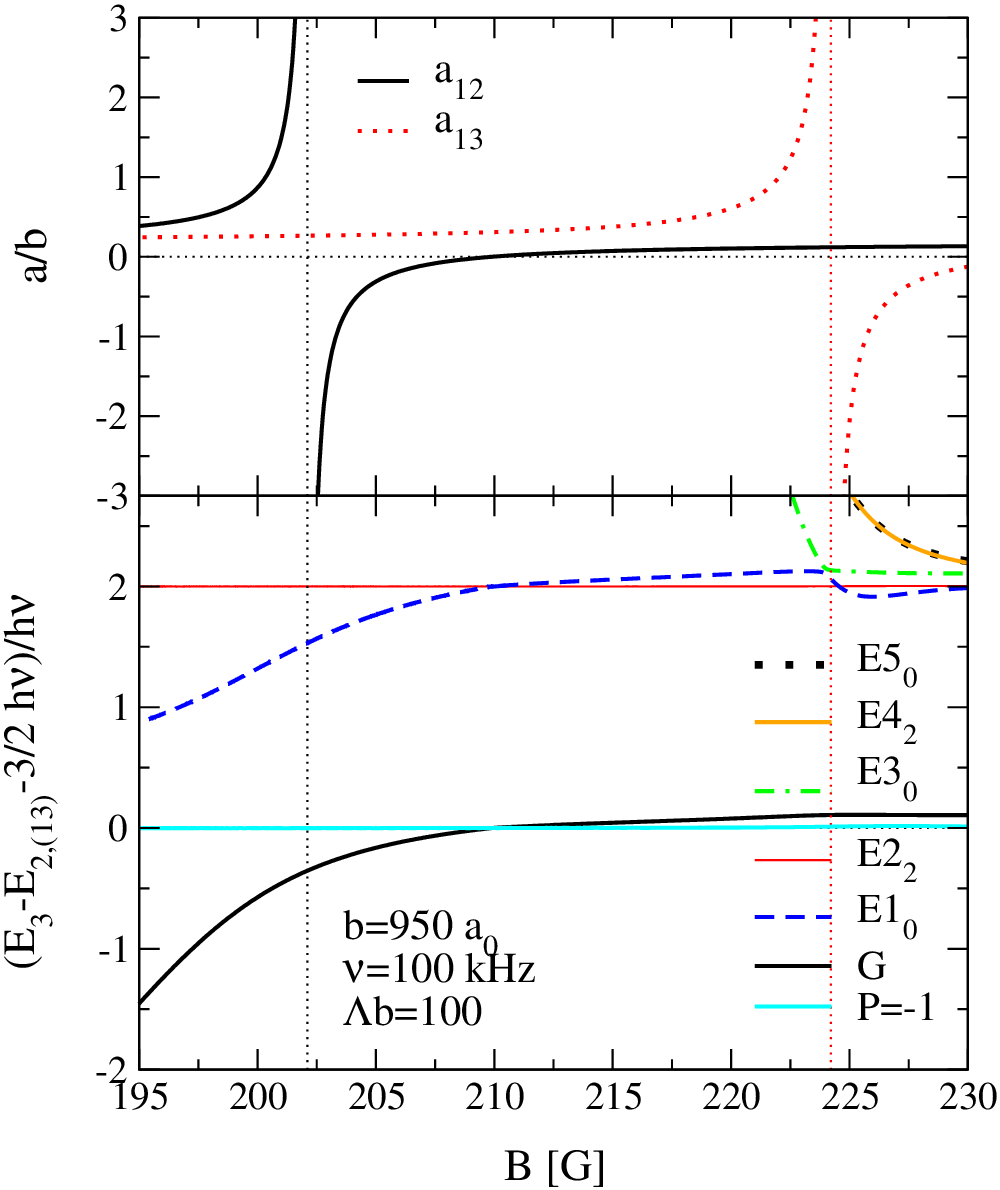}
\caption{Top: Feshbach resonances between the three
lowest hyperfine states of $^{40}$K atoms~[10-12].
Bottom: The spectrum for the above $^{40}$K states
in a single well of an optical lattice with $\nu = 100 \khz$ 
versus magnetic field. These BH results are for $a_{23}=0$
and $\la b = 100$. The three-body energy $E_3$ is measured 
from the three-body dissociation threshold given by the 
ground-state energy of pair $13$ ($E_{2,(13)}$) and the 
additional particle in $E_1 
= 3/2 \hbar \omega$ ($5/2 \hbar \omega$ for P=-1).
The state labeled G is the three-body ground state and En$_L$
is as in Fig.~\ref{Li6}.
\label{K40}}
\end{center}
\end{figure}

{\it Three-fermion problems.--} Across a Feshbach resonance the
dependence of the scattering length on the magnetic field $B$ is 
given by $a(B) = a_{\rm bg} ( 1 - \Delta/[B - \overline{B}])$ 
where $\overline{B}$ and $\Delta$ are the position 
and width of the resonance
and $a_{\rm bg}$ denotes the background scattering length. For $^6$Li, 
the three trapped hyperfine states are the lowest magnetic sub-states: 
$| 1 \rangle = | F, m_F \rangle = | 1/2, 1/2 \rangle$, $| 2 \rangle 
= | 1/2, -1/2 \rangle$ and $| 3 \rangle = | 3/2, -3/2 \rangle$, with 
distinct Feshbach resonances, as shown in Fig.~\ref{Li6}. The resonance 
positions are $\overline{B}_{12}=83.41 \mt$, $\overline{B}_{13}=69.04 
\mt$ and $\overline{B}_{23}=81.12 \mt$~\cite{Li}, and we use the
Feshbach parameters plus leading-order correction 
determined in~\cite{Li}. The relevant hyperfine states 
for $^{40}$K are $| 1 \rangle = | F, m_F \rangle = | 9/2, -9/2 \rangle$, 
$| 2 \rangle = | 9/2, -7/2 \rangle$ and $| 3 \rangle = | 9/2, -5/2 
\rangle$. As shown in Fig.~\ref{K40}, nearby Feshbach resonances are
present between the states $12$ and $13$, with Feshbach parameters: 
$\overline{B}_{12} = 202.10 \g$~\cite{RGJ}, %\pm0.07
$\Delta_{12} = 7.8 \g$~\cite{RTBJ}, %\pm0.6
$\overline{B}_{13} = 224.21 \g$, %\pm0.05
$\Delta_{13} = 9.7 \g$~\cite{RJ} %\pm0.6
and background scattering length $a_{\rm bg} \approx 174 \, a_0$~\cite{RJ}.

We work with effective field-theory contact interactions regulated 
by separable cutoff functions~\cite{BvK}
\be
V(p',p) = \frac{4\pi \hbar^2}{m} \, g(B,\la) \, 
e^{-(p'^2+p^2)/\la^2} \,,
\label{Vsep}
\ee
where $p$ and $p'$ denote incoming and outgoing relative momenta,
and the coupling $g(B,\la)$ is determined from the scattering 
length through
$g(B,\la) = a(B)/(1 - \la a(B)/\sqrt{2\pi})$.
If $a$ is weak ($a \sim R$, where $R$ is the 
range of the interaction), it is possible to choose the cutoff 
in a wide range with $|\la a| \ll 1$, and one recovers the
standard pseudo-potential for low momenta $V(0,0) = 4 \pi \hbar^2
a/m$. The cutoff generates an effective range $r_{\rm e} \sim 
1/\la$ and higher-order terms, which we render small with large
cutoffs. In addition we can vary $\la$. This probes
neglected effective range effects and sensitivity to atomic details.

For the separable interaction, Eq.~(\ref{Vsep}), the two-body 
problem in a harmonic oscillator potential can be solved exactly. 
Following Busch {\it et al.}~\cite{Busch}, the intrinsic energy 
$E = \epsilon \, \hbar \omega$ is given by
\be
\frac{_2F_1\bigl(\frac{3}{2},\frac{3}{4}-\frac{\epsilon}{2};
\frac{7}{4}-\frac{\epsilon}{2};
\bigl[\frac{1-x^2}{1+x^2}\bigr]^2\bigr)}{
\bigl(1+\frac{1}{x^2}\bigr)^3 \bigl(\frac{\epsilon}{2}-\frac{3}{4}\bigr)}
+ x = \frac{\sqrt{2\pi}}{a/b} \,,
\label{elam}
\ee
where $_2F_1$ is the hypergeometric function, $x=\la b$ and $b = 
\sqrt{\hbar/m \omega}$ is the
oscillator length. For large cutoffs $x \to \infty$, we recover
the result of Busch {\it et al.}, $\sqrt{2} \, \Gamma(\frac{3}{4}
-\frac{\epsilon}{2}) / \Gamma(\frac{1}{4}-\frac{\epsilon}{2})=
b/a$~\cite{Busch}. The energy from Eq.~(\ref{elam}) is within $3 
\%$ (or $7 \%$) of the latter for $\la b = 100$ and all 
scattering lengths except the tight-binding region $0 < a/b 
\leqslant 1$ (or $0.5$). Typical well frequencies in optical 
lattices are $\nu \sim 100 \khz$. Consequently, the oscillator 
length $b \sim 1000 \, a_0$ ($a_0$ denotes the Bohr radius)
is large compared to the range of atomic interactions
$R \sim 10 \, a_0$. Finally, 
the measured ground-state energies of two particles in a single 
well of an optical lattice agree very well with this 
result~\cite{Michael} even down to $|a/b| \approx 1$.

{\it Bloch-Horowitz method.--} The spectrum for the intrinsic 
energy of three hyperfine states in a harmonic well is determined 
from the three-body Hamiltonian
\begin{multline}
H = H_0 + V = \sum\limits_{i=1}^3 \, H_0({\bf r}_i) 
- H_{0,{\rm cm}}({\bf R}) \\
+ V_{12}({\bf r}_1,{\bf r}_2) + V_{13}({\bf r}_1,{\bf r}_3) 
+ V_{23}({\bf r}_2,{\bf r}_3) \,,
\end{multline}
where the noninteracting part is given by $H_0({\bf r}) = 
-\hbar^2 \nabla_{\bf r}^2/2m + m \omega^2 r^2/2$ and ${\bf R}
=({\bf r}_1+{\bf r}_2+{\bf r}_3)/3$ denotes the center-of-mass 
(cm) coordinate ($H_{0,{\rm cm}}$ uses mass $3m$). Within
Jacobi coordinates, the eigenstates of $H_0$ are characterized
by
\be
H_0 \, | 12, 3 \rangle = (N+3) \, \hbar \omega \, | 12, 3 \rangle \,.
\ee
Here, $12 \equiv n_{12}$, $l_{12}$, $m_{12}$ are the radial, angular and
magnetic quantum numbers of the pair $12$, and $3 \equiv n_3$, $l_3$, 
$m_3$ refer to the quantum numbers of the third particle with 
respect to the cm of pair $12$. We classify these
noninteracting states $| N_i \rangle$ according to their 
principal quantum number $N = 2 n_{12} + l_{12} + 2 n_3 + 
l_3$, where the subindex $i$ denotes all possible states for
fixed total $N$.

We solve the three-body problem $(H_0+V)|\psi(E)\rangle =
E |\psi(E)\rangle$ using the Bloch-Horowitz (BH)
approach~\cite{BHmethod,HSL}. The BH method diagonalizes
an effective Hamiltonian in a truncated space of $P = \sum_{N_i
\leqslant \nmax} |N_i \rangle\langle N_i|$ low-energy
excitations, such that the low-lying spectrum is exactly
reproduced. Inserting $1=P+Q$, we obtain the projections
of the three-body Schr\"odinger equation:
\begin{align}
P (H_0+V) (P+Q) |\psi(E)\rangle &= E \, P |\psi(E)\rangle \,, 
\label{Peq} \\[1mm]
Q (H_0+V) (P+Q) |\psi(E)\rangle &= E \, Q |\psi(E)\rangle \,.
\label{Qeq}
\end{align}
Since $[P,H_0]=[Q,H_0]=0$ and $PQ=0$, we can solve Eq.~(\ref{Qeq})
for $Q |\psi(E)\rangle = (E-H_0)^{-1} Q V |\psi(E)\rangle$ and
insert the latter into Eq.~(\ref{Peq}). This leads to the
equivalent problem in the truncated space,
\be
P \biggl( H_0 + V + V \frac{Q}{E-H_0} V \biggr) P |\psi(E)\rangle
= E \, P |\psi(E)\rangle \,,
\label{Heff}
\ee
where the effective Hamiltonian $H_{\rm eff}(E)$ (given by the
operator in parentheses in Eq.~(\ref{Heff})) depends self-consistently
on the exact energy and exactly reproduces the low-lying spectrum,
as long as the eigenstate has overlap with the truncated space
$P |\psi(E)\rangle \neq 0$. Finally, we use a Faddeev decomposition
$|\psi(E)\rangle = (1 + P_{12} P_{13} + P_{12} P_{23}) |\psi(E)
\rangle_{12}$ to construct $H_{\rm eff}(E)$ ($P_{ij}$ is the
permutation operator).

The good quantum numbers of the interacting eigenstates are 
parity $P$, total angular momentum $L$ and projection $L_z$:
$|\psi(E)\rangle = |E;P,L,L_z\rangle$. Here we solve the
BH Eq.~(\ref{Heff}) in an uncoupled basis for
$L_z=0$ ($m_{12}=m_3=0$). Since $|E;P,L_z\rangle = \sum_L 
C_L(E;P,L_z) \, |E;P,L,L_z\rangle$, the resulting spectra
automatically contain states with all possible angular
momentum quantum numbers. We will use the BH overlap 
condition to identify their angular momenta. The
BH method has been used to calculate the ground-state
properties of light nuclei~\cite{HSL}, and as a check,
we have reproduced the results of Stoll and K\"ohler
for three identical bosons~\cite{SK}. For the separable
interaction, Eq.~(\ref{Vsep}), it is possible to
calculate the necessary BH two-body matrix elements
analytically. In addition, we have found it sufficient
to keep $l_{12}, l_3 \leqslant 3$.

\begin{figure}[t]
\begin{center}
\includegraphics[scale=0.44,clip=]{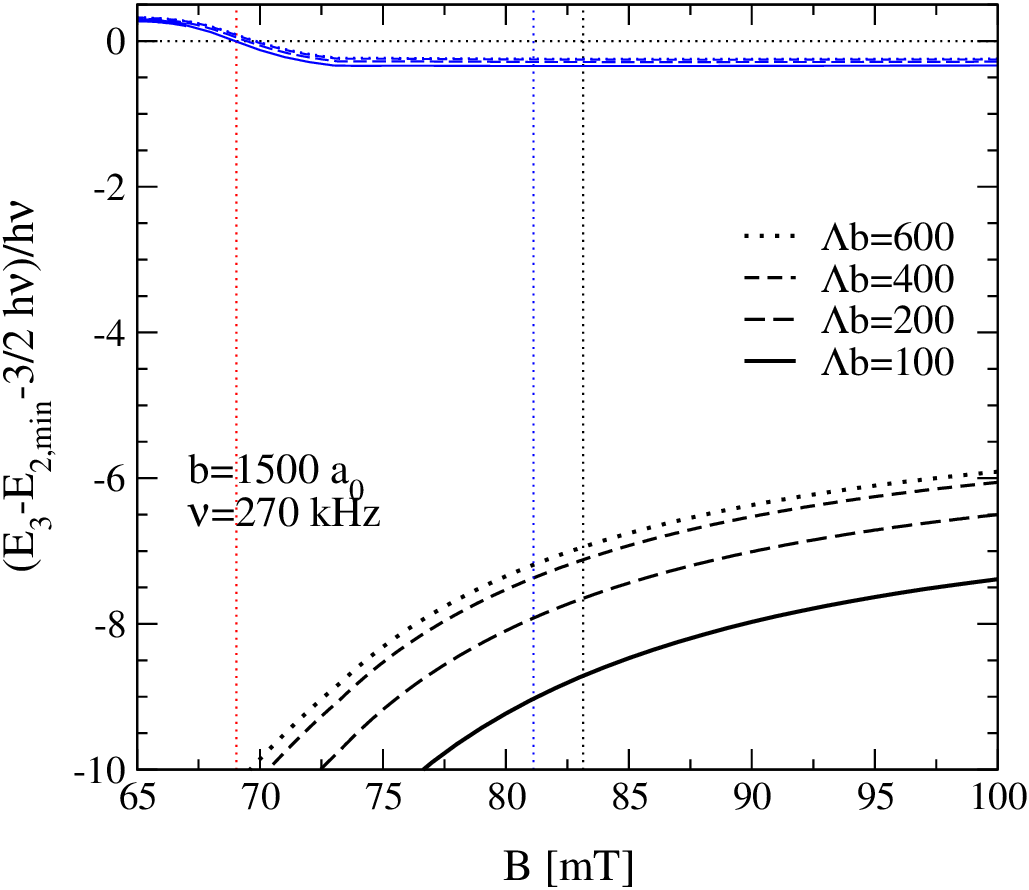}
\caption{The cutoff dependence of the Borromean and the first 
excited state of three $^6$Li atoms versus magnetic field.
\label{Li6_lb}}
\end{center}
\end{figure}

{\it Results.-- } The magnetic field dependence of the spectrum 
for three $^6$Li atoms in an optical lattice with $\nu = 270 
\khz$ is shown in Fig.~\ref{Li6}. The BH results are independent 
of $N_{\rm max}$. In particular, all states are present in the
lowest $N_{\rm max}=0$ calculation (with $l_{12}=l_3=0$), and 
thus all positive parity states shown have angular momentum $L=0$. 
The lowest negative parity state has $L=1$.

We find a deeply-bound Borromean state B that exists on the negative
scattering length side and extends across the Feshbach resonances.
Note that there are many very deeply-bound two-body states present.
This state can be viewed as a collective state within a schematic
model~\cite{Gerry}. For $B > 75 \mt$, the excited and negative
parity states depend very weakly on the magnetic field in 
Fig.~\ref{Li6}, since the Feshbach resonances of $a_{12}$ and 
$a_{23}$ are very close and here this two-body energy is subtracted.
The first excited state E1$_0$ is adiabatically connected to the
noninteracting $N=0$ state at high magnetic fields. Similarly,
the states E2$_0$ and E3$_0$ connect to the two $N=2$ states
with $l_{12}=l_3=0$, where the other three states of the
noninteracting $N=2$, $L_z=0$ multiplet are higher in
energy since they are less sensitive to S-wave interactions.

\begin{figure}[t]
\begin{center}
\includegraphics[scale=0.44,clip=]{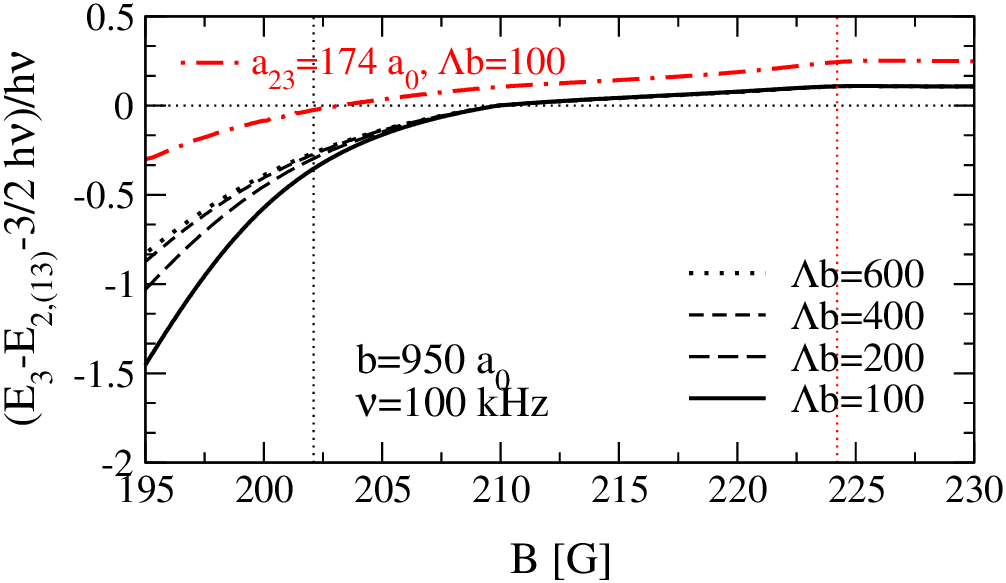}
\caption{The cutoff dependence of the ground state of three 
$^{40}$K atoms versus magnetic field.
\label{K40_lb}}
\end{center}
\end{figure}

In Fig.~\ref{K40} we show the spectrum for three $^{40}$K
atoms in an optical lattice with $\nu = 100 \khz$ versus
magnetic field. In this case, there are two Feshbach
resonances between pairs $12$ and $13$, and we have taken the
third pair to be noninteracting $a_{23}=0$ in this calculation.
We find that a bound three-body state only develops 
towards the positive scattering-length side of both resonances.
The three-body state is bound by the doubly-interacting
particle 1. The qualitative features of the spectrum do not 
depend on $a_{23}$ for $|a_{23}/a_0| \lesssim 100$. For
instance, with a repulsive $a_{23}=a_{\rm bg} \approx 174 
\, a_0$, the spectrum is moved up and the ground state
becomes bound only for lower magnetic fields (see
Fig.~\ref{K40_lb}, note that the precise value of
$a_{23}$ has not been calculated).

Two of the states of Fig.~\ref{K40} (E2$_2$ and E4$_2$)
are not present in a $N_{\rm max}=0$ calculation, but
exist for all larger $N_{\rm max} \geqslant 2$, and thus
have angular momentum $L=2$. At high magnetic fields,
we recover the five states adiabatically connected to
$N=2$ (note $l_{13}=l_2=1$ can couple to $L=0$, and also
the avoided level crossing for two of the $L=0$ states).
For $B_0=209.9 \g$, we have $a_{12}(B_0)=0$, and the
only interaction is for pair $13$. Therefore, the low-lying 
states are $E_3 - E_{2,(13)}-3/2\hbar\omega
=  2n_2+l_2$ for the positive parity state (the excitation
of the cluster $13$ comes higher in energy; and $2n_2+l_2-1$ 
with $-5/2\hbar\omega$ for P=-1), in agreement with
Fig.~\ref{K40}. The states E2$_2$
($l_2=2$) and P=-1 ($l_2=1$) follow only the two-body
energy of the right Feshbach resonance. Finally
for $B > B_0$, the interaction between $12$ becomes weakly 
repulsive, which
requires $\la_{12} a < \sqrt{2 \pi}$. 
Our $^{40}$K results are for $\la b = 100$, except $\la_{12}
a_{\rm bg} = 1$ for $B > B_0$. For $\la_{12} a < \sqrt{2 
\pi}$, we find a very weak cutoff dependence from the 
repulsive part of the $12$ interaction.

We can vary the cutoff and thus probe the
dependence of our results to the effects of an effective
range and many-body interactions~\cite{BvK}. 
In Figs.~\ref{Li6_lb} and~\ref{K40_lb}, we show
the cutoff dependence of the Borromean and first excited
states for $^6$Li, and the ground state of the $^{40}$K 
three-body problem. While this excited state (and
all others) are well
converged, the ground state energies converge slower
and show a sizeable dependence on $\la b$, as the binding
energy increases. Therefore, these states 
become sensitive to further atomic
details, such as the effective range. Our results also
indicate that there is no limit cycle in a harmonic oscillator
potential with $b/R \lesssim 100$, and thus the power-counting 
of three-body interactions in the corresponding effective field 
theory (EFT) must change compared to free space. This may be 
important for a pionless EFT~\cite{BvK} for nuclei in an
oscillator basis.

{\it Conclusions.-- } Optical lattices open a frontier 
to controlled strong-interaction few-body 
physics, in addition to simulating condensed matter models.
In this Letter, we investigated three-fermion problems in 
optical lattices for $^6$Li and $^{40}$K atoms using the BH method.
For $^6$Li atoms, we find a Borromean state on the negative
scattering-length side that extends across the Feshbach resonances.
In contrast, for $^{40}$K atoms, one of the pairs interacts
non-resonantly at the relevant magnetic fields, and a three-body
state, bound by the doubly-interacting particle, develops towards 
the positive scattering-length side.
While the quantitative results of the ground states are
somewhat sensitive to atomic details, these features are
independent thereof and also independent of the precise
oscillator frequency. The three-fermion spectra are very
rich, and we have identified the nature and angular momenta
of all low-lying states. We predict a Borromean
state in optical lattices under the conditions of overlapping 
or close Feshbach resonances for all pairs and attractive 
background scattering lengths.

\acknowledgments 
We especially thank M. K\"ohl for many discussions. We also
thank A. Bulgac, M. Forbes, C. Forss\'en, R. Furnstahl, V. 
Gueorguiev and H.-W. Hammer.
This work was performed under the auspices of the US DOE by the 
University of California, Lawrence Livermore National Laboratory 
under Contract~No. W-7405-Eng-48, and supported by the US DOE 
under Grant No.~DE-FG02-97ER41014.


\begin{thebibliography}{99}
\bibitem{Jaksch} D. Jaksch, {\it et al.}, Phys. Rev. Lett. 
\textbf{89}, 040402 (2002).
\bibitem{BH} E. Braaten and H.-W. Hammer, Phys. Rep. \textbf{428},
259 (2006).
\bibitem{Michael} T. St\"oferle, {\it et al.}, Phys. Rev. Lett. 
\textbf{96}, 030401 (2006).
\bibitem{Efimov} V.N. Efimov, Phys. Lett. \textbf{33B}, 563 (1970);
V.N. Efimov, Sov. J. Nucl. Phys. \textbf{12}, 589 (1971).
\bibitem{Kraemer}  T. Kraemer, {\it et al.}, Nature \textbf{440}, 
315 (2006). 
\bibitem{SK} M. Stoll and T. K\"ohler, Phys. Rev. \textbf{A72}, 
022714 (2005).
\bibitem{JHP} S. Jonsell, H. Heiselberg and C.J. Pethick, Phys. Rev. 
Lett. \textbf{89}, 250401 (2002).
\bibitem{Zhukov} M.V. Zhukov, {\it et al.}, Phys. Rep. \textbf{231},
151 (1991).
\bibitem{Li} M. Bartenstein, {\it et al.}, Phys. Rev. Lett. 
\textbf{94}, 103201 (2005).
\bibitem{RGJ} C.A. Regal, M. Greiner and D.S. Jin, Phys. Rev. Lett.
\textbf{92}, 040403 (2004); Phys. Rev. Lett. \textbf{92}, 083201 (2004).
\bibitem{RJ} C.A. Regal and D.S. Jin, Phys. Rev. Lett. \textbf{90}, 
230404 (2003).
\bibitem{RTBJ} C.A. Regal, {\it et al.}, Phys. Rev. Lett. 
\textbf{90}, 053201 (2003).
\bibitem{BvK} P.F. Bedaque and U. van Kolck, Ann. Rev. Nucl. Part.
Sci. \textbf{52} (2002) 339.
\bibitem{Busch} T. Busch, {\it et al.}, Found. Phys. \textbf{28}, 
549 (1998).
\bibitem{BHmethod} C. Bloch and J. Horowitz, Nucl. Phys. \textbf{8}, 
91 (1958).
\bibitem{HSL} W.C. Haxton and C.-L. Song, Phys. Rev. Lett. \textbf{84}, 
5484 (2000); W.C. Haxton and T. Luu, Phys. Rev. Lett. \textbf{89},
182503 (2002).
\bibitem{Gerry} G.E. Brown and M. Bolsterli, Phys. Rev. Lett.
\textbf{3}, 472 (1959).

\end{thebibliography}
\end{document}